\documentclass[12pt]{article}
\usepackage{amsmath}
\usepackage{amssymb}
\usepackage{amsfonts}
\usepackage{latexsym}
\usepackage{color}
\usepackage{graphicx}

\catcode `\@=11 \@addtoreset{equation}{section}

\catcode `\@=12

\newcommand{\be}{\begin{equation}}
\newcommand{\en}{\end{equation}}
\newcommand{\bea}{\begin{eqnarray}}
\newcommand{\ena}{\end{eqnarray}}
\newcommand{\beano}{\begin{eqnarray*}}
\newcommand{\enano}{\end{eqnarray*}}
\newcommand{\bee}{\begin{enumerate}}
\newcommand{\ene}{\end{enumerate}}

\newcommand{\Hil}{{\cal H}}

\newcommand{\Id}{1\!\!1}
\newcommand{\F}{{\cal F}}
\newcommand{\Lc}{{\cal L}}
\newcommand{\Sc}{{\cal S}}
\newcommand{\D}{{\cal D}}
\newcommand{\C}{{\cal C}}

\newcommand{\E}{{\cal E}}
\newcommand{\Nc}{{\cal N}}
\newcommand{\M}{{\cal M}}

\newcommand{\1}{1 \!\!\! 1}
\newtheorem{thm}{Theorem}

\newtheorem{prop}[thm]{Proposition}

\begin{document}

\thispagestyle{empty}

\vspace*{1cm}

\begin{center}
{\Large \bf An invariant analytic orthonormalization procedure with  applications}   \vspace{2cm}\\

{\large M.R. Abdollahpour}\\
  Dept. of Mathematics,
University of Tabriz, Tabriz, Iran\\
e-mail: m.abdollah@tabrizu.ac.ir \vspace{5mm}\\
{\large F. Bagarello}\\
  Dipartimento di Metodi e Modelli Matematici,
Fac. Ingegneria, Universit\`a di Palermo, I-90128  Palermo, Italy\\
e-mail: bagarell@unipa.it
\vspace{5mm}\\
 {\large S. Triolo}\\
  Dipartimento di Matematica ed Applicazioni,
 Universit\`a di Palermo, I-90128  Palermo, Italy\\
e-mail: salvo@math.unipa.it
\end{center}

\vspace*{2cm}

\begin{abstract}
\noindent We apply the orthonormalization procedure previously
introduced by two of us and adopted in connection with coherent
states to Gabor frames and other examples. For instance, for Gabor
frames we show how to construct $g(x)\in\Lc^2(\Bbb{R})$ in such a
way the functions $g_{\underline n}(x)=e^{ian_1x}g(x+an_2)$,
$\underline n\in\Bbb{Z}^2$ and $a$ some positive real number, are
mutually orthogonal. We discuss in some details the role of the
lattice naturally associated to the procedure in this analysis.

\end{abstract}

\vspace{2cm}


\vfill

\newpage

\section{Introduction and mathematical results}

In the mathematical and physical literature many examples of
complete sets of vectors in a given Hilbert space $\Hil$ are
constructed starting from a single normalized element $f_0\in\Hil$,
acting on this vector several time with a given set of unitary
operators. This is exactly what happens, for instance, for coherent
states and for wavelets, as well as for Shannon systems and Gabor
frames. All these examples can be considered as particular cases of
a general procedure in which a certain set of vectors is constructed
acting on a fixed element of $\Hil$, $f_0$, with a certain set of
unitary operators, $A_1,\ldots,A_N$:
$f_{k_1,\dots,k_N}:=A_1^{k_1}\cdots A_N^{k_N}f_0$, $k_j\in\Bbb{Z}$
for all $j=1,2,\ldots,n$. These vectors may or may not be
orthogonal: we consider here the problem of {\em orthonormalizing
this set}, i.e. the problem of producing a new set of vectors which
share with the original one most of its features (we will be more
precise in the following) and, moreover, are also orthonormal. In
\cite{bagtri} we have constructed a general strategy for achieving
this aim and we have applied it to coherent states, producing a new
set of square-integrable functions which have many of the properties
of coherent state and, moreover, are mutually orthogonal. These
results will be reviewed in Section II, where we also  describe the
method.

In Section III we apply the strategy to Shannon frame, while Section
IV is dedicated to a deep analysis of Gabor frames. In particular we
will discuss the role of what we call {\em the lattice spacing} in
the orthonormalization procedure.

Section V is devoted to some examples, which we will discuss in
details, and to our conclusions.

\section{Stating the problem and first results}

Let $\Hil$ be a Hilbert space, $f_0\in\Hil$ a fixed element of the
space and $A_1,\ldots,A_N$, $N$ given unitary operators:
$A_j^{-1}=A_j^\dagger$, $j=1,2,\ldots,N$. Let $\Hil_N$ be the
closure of the linear span of the set
\be\Nc_N=\{f_{k_1,\ldots,k_N}:=A_1^{k_1}\cdots A_N^{k_N}f_0,\,\,
k_1,\ldots k_N\in\Bbb{Z} \}\label{II1}\en which we assume to
consists of linearly independent vectors in order not to be trivial.
Of course the vectors $f_{k_1,\ldots,k_N}$ are  complete in $\Hil_N$
by definition and $\Hil_N$ could coincide or not with all of $\Hil$.
In general there is no reason why the vectors in $\Nc_N$ should be
mutually orthogonal. On the contrary, without a rather clever choice
of both $f_0$ and $A_1,\ldots,A_N$, it is very unlikely to obtain
such an o.n. set. Our aim is to discuss some general technique which
produces another normalized vector $\varphi\in\Hil_N$ such that the
set \be\M_N=\{\varphi_{k_1,\ldots,k_N}:=A_1^{k_1}\cdots
A_N^{k_N}\varphi,\,\, k_1,\ldots ,k_N\in\Bbb{Z} \}\label{II2}\en is
made of orthogonal vectors. Moreover, we would like this set to
share as much of the original features of $\Nc_N$ as possible. For
instance, if the set $\Nc_N$ is a set of coherent states, we have
shown in \cite{bagtri} that the new vectors
$\varphi_{k_1,\ldots,k_N}$ are, among the other features,
eigenstates of a (sort of) annihilation operator, give rise to a
resolution of the identity and they saturate the Heisenberg
uncertainty relation: so they appear, in some way, as a set of
generalized orthonormal coherent states.

The easiest situation,  $N=1$, goes like this: in this case we have
$\Nc_1 =\{f_{k}:=A^{k}\,f_0,\,\, k\in\Bbb{Z} \}$ with $<f_k,f_l>\neq
\delta_{k,l}$ (otherwise we have already solved the problem!). Since
$\Nc_1$ is complete in $\Hil_1$, any element in $\Hil_1$ can be
written in terms of the vectors of $\Nc_1$. Let $\varphi_0\in\Hil_1$
be the following linear combination
$\varphi_0=\sum_{k\in\Bbb{Z}}\,c_kf_k,$ and let us define more
vectors of $\Hil_1$ as
$\varphi_n=A^n\varphi_0=\sum_{k\in\Bbb{Z}}\,c_k\,f_{k+n}=Xf_{n},$
where we have introduced the  operator $
X=\sum_{k\in\Bbb{Z}}\,c_k\,A^k.$ The coefficients $c_k$ should be
fixed by the following orthogonalization requirement:
$<\varphi_n,\varphi_0>=\delta_{n,0}$, for $n\in\Bbb{Z}$. As we have
discussed in \cite{bagtri}, these expansions are, up to this stage,
purely formal. However, in many concrete relevant situations they
converge and, if this is so, we find that the coefficients are \be
c_l=\frac{1}{2\pi}\int_0^{2\pi}\,
\frac{e^{-ipl}\,dp}{\sqrt{\alpha(p)}}\label{II5bis}\en where the
function $\alpha(p)$ is defined as
$\alpha(p)=\sum_{l\in\Bbb{Z}}\,a_l\,e^{ipl}$, with
$a_j=<A^jf_0,f_0>$. A sufficient condition for $\alpha(p)$ to exist
is that  $\{a_j\}\in l^1(\Bbb{Z})$, which, in turns, is related to
the nature of both $A$ and $f_0$. We refer to \cite{bagtri} and to
the next Section for more results and examples concerning $N=1$.

\vspace{2mm}

Let us now take $N=2$. We quickly review here our previous results
on coherent states.

Let $\hat q$ and $\hat p$ be the position and momentum operators on
the Hilbert space $\Hil=\Lc^2(\Bbb{R})$, $[\hat q,\hat p]=i\Id$, and
let us now introduce the following unitary operators: \be
U(\underline n)=e^{ia(n_1\hat q-n_2\hat p)}, \quad D(\underline
n)=e^{z_{\underline{n}} b^\dagger-\overline{z}_{\underline n}
b},\quad T_1:=e^{ia\hat q},\quad T_2:=e^{-ia\hat p}.\label{31}\en
Here $a$ is a real constant satisfying $a^2=2\pi L$ for some
$L\in\Bbb{N}$, while $z_{\underline{n}}$ and $b$ are related to
${\underline{n}}=(n_1,n_2)$ and $\hat q$, $\hat p$ via the following
equalities: \be z_{\underline{n}}=\frac{a}{\sqrt{2}}(n_2+in_1),\quad
b=\frac{1}{\sqrt{2}}(\hat q+i\hat p). \label{32}\en With these
definitions it is clear that \be U(\underline n)=D(\underline
n)=(-1)^{Ln_1n_2}T_1^{n_1}T_2^{n_2}=(-1)^{Ln_1n_2}T_2^{n_2}T_1^{n_1},\label{33}\en
where we have also used the commutation rule $[T_1,T_2]=0$.

Let $\varphi_{\underline{0}}$ be the vacuum of $b$,
$b\varphi_{\underline{0}}=0$, and let us define the following {\em
coherent states}: \be
\varphi_{\underline{n}}^{(L)}:=T_1^{n_1}T_2^{n_2}\varphi_{\underline{0}}=T_2^{n_2}T_1^{n_1}
\varphi_{\underline{0}} =(-1)^{Ln_1n_2}U(\underline
n)\varphi_{\underline{0}}=(-1)^{Ln_1n_2}D(\underline
n)\varphi_{\underline{0}},\label{34}\en \cite{kl,aag}. It is  very
well known that the set of these vectors,
$\C^{(L)}=\{\varphi_{\underline{n}}^{(L)},\,\underline{n}\in{\Bbb{Z}}^2\}$,
satisfies a number of relevant properties\footnote{(a) $\C^{(L)}$ is
invariant under the action of $T_j^{n_j}$, for all $n_j$, $j=1,2$;
(b) each $\varphi_{\underline{n}}^{(L)}$ is an eigenstate of $b$:
$b\varphi_{\underline{n}}^{(L)}=z_{\underline{n}}\,\varphi_{\underline{n}}^{(L)}$;
(c) they satisfy the {\em resolution of the identity} on a certain
Hilbert space $h_L$, see (\ref{36}),
$\sum_{\underline{n}\in{\Bbb{Z}}^2}\,|\varphi_{\underline{n}}^{(L)}><\varphi_{\underline{n}}^{(L)}|=\Id_L$,
where $\Id_L$ is the identity on $h_L$; (d) They saturate the
Heisenberg uncertainty principle: if we call $(\Delta
X)^2=<X^2>-<X>^2$ for $X=\hat q,\hat p$, then $\Delta \hat q\,\Delta
\hat p=\frac{1}{2}$.}. However, it is also well known that they are
not mutually orthogonal. Indeed we have
$I_{\underline{n}}^{(L)}:=<\varphi_{\underline{n}}^{(L)},\varphi_{\underline{0}}>=(-1)^{Ln_1n_2}\,e^{-\frac{\pi}{2}L(n_1^2
+n_2^2)}$, which is only {\em nearly} zero if $L$ is large enough
and $(n_1,n_2)\neq (0,0)$.

Our aim is to construct a family of vectors $\E^{(L)}$ which shares
with $\C^{(L)}$ most of the above features and which, moreover, is
made of orthonormal vectors. We have shown in \cite{bagtri} that
this is possible, in suitable Hilbert spaces, if $L>1$, while
complications arise for $L=1$. Incidentally, we have shown that the
set $\C^{(L)}$ is complete in $\Hil$ if and only if $L=1$.
\vspace{2mm}

Let us now define, for each $L\geq 1$, the following set: \be
h_L:=\mbox{linear
span}\overline{\left\{\varphi_{\underline{n}}^{(L)},\,\underline{n}\in{\Bbb{Z}}^{
2}\right\}}^{\|.\|}. \label{36}\en We know that $h_1=\Hil$, while,
for $L>1$, $h_L\subset \Hil$. It is further clear that $h_L$ is an
Hilbert space for each $L$, since it is a closed subspace of $\Hil$.

Now we define \be
\Psi_{\underline{n}}^{(L)}:=\sum_{\underline{k}\in{\Bbb{Z}}^2}\,c_{\underline{k}}^{(L)}\,
\varphi_{\underline{k}+\underline{n}}^{(L)} \label{37} \en Of course
this means that
$\Psi_{\underline{0}}^{(L)}:=\sum_{\underline{k}\in{\Bbb{Z}}^2}\,c_{\underline{k}}^{(L)}\,
\varphi_{\underline{k}}^{(L)}$ and, because of the commutativity of
$T_1$ and $T_2$, we also have \be
\Psi_{\underline{n}}^{(L)}=T_1^{n_1}\,T_2^{n_2}\Psi_{\underline{0}}^{(L)}.
\label{38}\en Therefore the new set constructed in this way,
$\E^{(L)}:=\{\Psi_{\underline{n}}^{(L)},\,\underline{n}\in{\Bbb{Z}}^2\}$,
is invariant under the action of $T_1$ and $T_2$, exactly as the set
$\C^{(L)}$, independently of the choice of the coefficients of the
expansion $c_{\underline{k}}^{(L)}$.  These coefficients, however,
must not be chosen freely: they are fixed  requiring that the
vectors in the set $\E^{(L)}$ are orthonormal:
$<\Psi_{\underline{n}}^{(L)},\Psi_{\underline{s}}^{(L)}>=\delta_{\underline{n},
\underline{s} }$. This will fix (not uniquely!) the value of the
$c_{\underline{s}}^{(L)}$'s, with a procedure which extends formula
(\ref{II5bis}). These coefficients, if they exist, can be found as
follows: \be
c_{\underline{k}}^{(L)}=\frac{1}{(2\pi)^2}\,\int_0^{2\pi}\int_0^{2\pi}\,\frac{e^{-i\underline{P}
\cdot\underline{k}}}{\sqrt{F_L(\underline{P})}}\,d\underline{P},\label{314}\en
where
$F_L(\underline{P})=\sum_{\underline{m}\in{\Bbb{Z}}^2}\,(-1)^{L\,m_1\,m_2}\,e^{-\frac{\pi}{2}\,L(m_1^2
+m_2^2)}\,e^{i\underline{P}\cdot\underline{m}}$.

As already mentioned, in \cite{bagtri} we have seen that $L=1$ and
$L>1$ are really different situations: if $L>1$ the above formulas
are well defined and produce a set $\E^{(L)}$ which has the same
properties of the vectors in $\C^{(L)}$ and, moreover, is made of
orthonormal vectors. On the contrary, if $L=1$, the procedure does
not work: the series are not convergent and $F_L(\underline P)$ has
a zero in $[0,2\pi[\times[0,2\pi[$. This is not related to our
method but it is rather a well known feature of coherent states:
indeed it is a standard result in functional analysis that in $\Hil$
no orthonormal set of coherent states can be constructed, \cite{kl}!

\section{Frames of translated}

We begin this section by recalling few known results concerning
frames of translated. After that, we will use our method to
construct an o.n. set of translated.

Assume that $\phi\in \Lc^{2}(\mathbb{R})$ and consider the set of
the form $\F=\{\phi(\cdot-k)\}_{k\in Z}=\{\hat T_{k}\phi\}$ where
$\hat T_{k}$ is related to the translation operator in (\ref{31}) as
follows: $\hat T_k:=T_2^k$, with $a=1$. Christensen, Deng, and Heil
in \cite{cdh} by using the Beurling densities proved that $\F$
cannot be complete in all of $\Lc^2(\Bbb{R})$ and that it can, at
most, be a frame for a proper subspace of $\Lc^{2}(\mathbb{R}).$
This is in line with what we have discussed in the previous section:
indeed the closure of the set $\Nc_1$ was not required to be all of
$\Hil$. What our strategy can produce here is, at most, an o.n. set
spanning the same Hilbert space which is spanned by the set $\F$.
\par Let us consider the function
\begin{equation}\label{phi}
\Phi:\mathbb{R}\rightarrow \mathbb{R},\qquad \Phi(p)=\sum_{k\in
Z}|\hat{\phi}(p+k)|^{2}
\end{equation}
 where $\hat{\phi}$ is the Fourier
transform of $\phi$,
$\hat\phi(p)=\frac{1}{\sqrt{2\pi}}\int_{\Bbb{R}}\phi(x)\,e^{-ipx}\,dx$.
The function $\Phi$  is 1-periodic and belongs to $\Lc^{1}([0,1])$.
This function can be used to check wether the set $\F$ is or not a
frame of translates. Indeed, see for instance \cite{ole} and
references therein, the following proposition holds true:
\begin{prop}
Let $\phi\in \Lc^{2}(\mathbb{R})$ then for $A,B>0$ the set $\F$ is
an $(A,B)$-frame  if and only if
$$A\leq\Phi(p)\leq B$$
a.e. in $p\in [0,1]\setminus\C$, where $\C=\{p\in[0,1]:\,
\Phi(p)=0\}$.
\end{prop}

\vspace{2mm}

{\bf Remark:} we want to stress  that we are here adopting a
notation which is slightly different when compared with \cite{ole}:
in particular we still call {\em frame} a certain set even if it
only spans a subset of the original Hilbert space, at least when
this aspect is clear and does not lead to confusion. In other words,
whenever the situation is clear from the context, we will say that a
certain set of translated is a frame even if it does not spans all
of $\Hil$.

\vspace{2mm}

Other results on frames of translates are discussed in many details
in \cite{ole}, to which we refer for more details.

\vspace{2mm}

Let us now apply our procedure for $N=1$ to a set of translated. For
that we suppose that an (A,B)-frame
$\Nc_1=\{\varphi_{j}\in\Hil\}_{j\in \mathbb{Z}}$ is generated by a
single fixed vector $\varphi$ and a single unitary operator $T\equiv
T_2$:
$$\varphi_{j}=T^{j}\varphi,\; j\in\mathbb{Z}.$$  So there exist $A,B>0$ such that
$$A\|f\|^{2}\leq \sum_{j\in \mathbb{Z}}|\langle T^j\varphi
,f\rangle|^{2}\leq B\|f\|^{2},\quad\forall f\in h,$$ where $h$ is
the subspace of $\Hil$ spanned by the vectors of $\Nc$. The
construction of the o.n. set goes as in the first part of the
previous section. Therefore we can define the coefficients $a_j$ as
usual, $a_j=<T^j\varphi,\varphi>$, and from these we find the
function $\alpha(p)$ and finally the coefficients $c_k$'s which
produce the vector $\varphi_0=\sum_{k\in\Bbb{Z}}\,c_kf_k$ as in
(\ref{II5bis}).

An example of this construction was already essentially discussed in
\cite{bagtri}. We consider this same example here from a slightly
different perspective.

{\bf Example:--} Let $g(x)=\chi_{[0,a[}(x)$ be the characteristic
function in the interval $[0,a[$, with $1<a<2$. Of course we have
$$\Nc_1=\{g_{n}(x):=T^{n}g(x)=\chi_{[n,n+a[}(x),\,\,
n\in\Bbb{Z} \}.$$ The overlap coefficients $a_j$ can
  be written as
$a_j=a\,\delta_{j,0}+(a-1)\left(\delta_{j,-1}+\delta_{j,1}\right)$,
so that $\alpha(p)=a+2(a-1)\cos(p)$. This is a nonnegative, real and
$2\pi$-periodic function and is never zero in $[0,2\pi[$ since it
has a minimum in $p=\pi$ and $\alpha(\pi)=2-a>0$. If we fix, as an
example, $a=\frac{3}{2}$, we can compute analytically
$\sum_{l\in{\Bbb{Z}}}\,|c_l|^2=\frac{1}{2\pi}\int_0^{2\pi}\frac{dp}{\alpha(p)}=\frac{2}{\sqrt{5}}$.
We refer to \cite{bagtri} for further considerations on this
example. Here we just want to comment that the function
$\Phi(p)=\sum_{k\in Z}|\hat g(p+k)|^{2}$ in (\ref{phi}) can be
written as $\Phi(p)=a+2(a-1)\cos(2\pi p)$, and this satisfies the
inequality $2-a<\Phi(p)<3a-2$. Therefore, using Proposition 1 above,
we conclude that the set $\{g_n(x), \,n\in\Bbb{Z}\}$ is a
$(2-a,3a-2)$-frame (which, of course, spans a proper subspace of
$\Lc^2(\Bbb{R})$). So we have a frame which, by means of our
strategy, produces an o.n. set which spans the same Hilbert subspace
of $\Lc^2(\Bbb{R})$ and which is stable under translations.

\vspace{2mm}

We would also like to mention that a different orthonormalization
procedure is well known to people working on wavelets since it may
be useful in the very first steps of a multi-resolution analysis to
construct an orthonormal basis of a certain subspace of
$\Lc^2(\Bbb{R})$, $V_0$, starting from a given Riesz basis of $V_0$.
This technique, which is reviewed in \cite{dau}, is mainly based on
the following fact: in $\varphi(x)$, together with its integer
translated $\varphi(x+n)$, $n\in\Bbb{Z}$ is a Riesz basis in
$V_0\subset\Lc^2(\Bbb{R})$, then the inverse Fourier transform
$\tilde\varphi(x)$ of the function
$\frac{1}{\sqrt{2\pi}}\left(\sum_{l\in\Bbb{Z}}\left|\hat\varphi(p+2\pi
l)\right|^2\right)^{    -1/2}\,\hat\varphi(p)$ is such that the set
$\{\tilde\varphi(x+n),\,n\in\Bbb{Z}\}$ is an orthonormal basis of
the same set $V_0$.

It is clear that this assumption exactly coincides with our original
hypothesis: any Riesz basis is a set of vectors $\{T_{k}\phi\}_{k\in
\mathbb{Z}}$ which are linearly independent and a system of
generator in the Hilbert space they generate, by definition.

It is also worth mentioning that the two techniques both rely on
Fourier analysis and, from this point of view, they look similar.
Nevertheless they are different and inequivalent since they produce,
in general, different results starting from the same {\em seed
function} $g(x)$ and, moreover, since the difficulties of the
computations are not necessarily comparable. For instance, if we
consider the previous example, our orthonormalization procedure
produces, at the end, the following function:
$\sum_n\left(\frac{1}{2\pi}\int_0^{2\pi}\frac{e^{-ipn}\,dp}{\sqrt{a+2(a-1)\cos(p)}}
\right)\,\chi_{[n,n+a[}(x)$, and the integrals can be easily
computed (Remember that we need to compute only very few of them,
because general arguments on Fourier series show that the
coefficients $c_n$ go to zero very fast with $n$, see below). A bit
more difficult, in our opinion, is to compute the function
$\tilde\varphi(x)$ using formula above. However both these
approaches in general can only be applied within an approximation
scheme. Finding an exact solution is, but for few examples, quite
difficult if not impossible. From this point of view our strategy
seems convenient with respect to the other one since it is very
natural to construct a perturbative approach. Indeed we only need to
check if the function $\alpha(p)$ has some zero in $[0,2\pi[$. If
this is not so, in fact, we know that the coefficients $c_m$ goes to
zero faster than any inverse power in $|m|$ and, therefore, the
series defining the new o.n. set converges very fast. This means
that only the first few $c_m$'s are required to get a reasonable
approximation of the solution we are looking for, and this can be
easily done numerically.

\vspace{2mm}

{\bf Remark:}  It is maybe not surprising to notice that both in our
and in the above approaches the following integrals play an
important role
$$a_{k}=\int_{-\infty}^{+\infty}\phi(t)\overline{\phi(t+k)}dt.$$
This is because $a_k$ is a measure of the  non orthogonality between
the {\em original} functions: the smaller is the difference of $a_k$
from $\delta_{k,0}$, the closer is the set $\{\phi(x+k),
\,k\in\Bbb{Z}\}$ to an orthonormal set.

\section{Gabor frames}

\subsection{What we get from $(k,q)$-representation}

We begin this section by recalling few results on the generalized
$(k,q)-$representation which we have introduced in \cite{bagtri},
and which extends analogous results discussed, for instance, in
\cite{zak2}.

Let  $T(a)=e^{i\hat p a}$ and $\tau(b)=e^{i\hat q b}$, with $ab=2\pi
L$, for some natural $L$. It is clear that for all possible
$L\in\Bbb{N}$ the two operators still commute: $[T(a),\tau(b)]=0$.
For each given $A>0$ let us define the set of (generalized)
functions \be\Phi_{k,q}^{(A,a)}(x)=
\sqrt{\frac{A}{2\pi}}\sum_{l\in\Bbb{Z}}\,e^{iklA}\,\delta(x-q-la),\label{28}\en
where
$(k,q)\in\Box^{(A)}:=\left[0,\frac{2\pi}{A}\right[\times[0,a[$. If
$\xi_x$ is the generalized eigenvector of the position operator
$\hat q$, $\hat q\xi_x=x\xi_x$, we write $\Phi_{k,q}^{(A,a)}(x)$ as
$\Phi_{k,q}^{(A,a)}(x)=<\xi_x,\Phi_{k,q}^{(A,a)}>$.

\begin{prop}
With the above definitions the following statements hold true:
\begin{enumerate}
\item \be T(a)\Phi_{k,q}^{(A,a)}(x)=e^{ikA}\Phi_{k,q}^{(A,a)}(x), \quad \tau(b)
\Phi_{k,q}^{(A,a)}(x)=e^{iqb}\Phi_{k,q}^{(A,a)}(x),\label{41}\en
\item
\be\int\int_{\Box^{(A)}}\overline{\Phi_{k,q}^{(A,a)}(x)}\,\Phi_{k,q}^{(A,a)}(x')\,dk\,dq=\delta(x-x'),\label{42}\en
\item
\be\int\int_{\Box^{(A)}}|\Phi_{k,q}^{(A,a)}><\Phi_{k,q}^{(A,a)}|=\Id,\label{43}\en
where the usual Dirac bra-ket notation has been adopted;
\item \be\int_{\Bbb{R}}\overline{\Phi_{k,q}^{(A,a)}(x)}\,\Phi_{k',q'}^{(A,a)}(x)
\,dx=\delta(k-k')\,\delta(q-q').\label{44}\en
\end{enumerate}
\end{prop}
The proof of these statements does not differ significantly from the
standard one, and will be omitted here. We just want to remark that,
for general $a$ and $a'$, we find that $T(a)\Phi_{k,q}^{(A,a')}\neq
e^{ikA}\Phi_{k,q}^{(A,a')}(x)$. In other words, in general
$\Phi_{k,q}^{(A,a')}(x)$ is not an eigenstate of $T(a)$ if $a\neq
a'$.

Also, it should be noticed that the value of the parameter $b$
entering in the definition of $\tau(b)$, is fixed by requiring that
$T$ and $\tau$ commute but play no role in the definition of the
lattice cell $\Box^{(A)}$, which on the other way is defined by an
extra positive parameter, $A$, which needs not to be related to $b$
itself. However, quite often in applications $A$ coincides with $a$
and with $b$.

\vspace{2mm}

We now use this generalized $(k,q)$-representation in connection
with Gabor frames and in the attempt of getting an o.n. set of
functions in $\Lc^2(\Bbb{R})$. The starting point is not very
different from that in Section II: let $\hat q$ and $\hat p$ be the
position and momentum operators on the Hilbert space
$\Hil=\Lc^2(\Bbb{R})$, $[\hat q,\hat p]=i\Id$, and let $U(\underline
n)=e^{ia(n_1\hat q-n_2\hat p)}$, $D(\underline
n)=e^{z_{\underline{n}} b^\dagger-\overline{z}_{\underline n} b}$,
$T_1:=e^{ia\hat q}$ and $T_2:=e^{-ia\hat p}$ as in (\ref{31}) As
before, $a$ is a real constant such that $a^2=2\pi L$ for some
$L\in\Bbb{N}$, while $z_{\underline{n}}$ and $b$ are related to
${\underline{n}}=(n_1,n_2)$ and $\hat q$, $\hat p$ as in (\ref{32}).
As in (\ref{33}) we have $U(\underline n)=D(\underline
n)=(-1)^{Ln_1n_2}T_1^{n_1}T_2^{n_2}=(-1)^{Ln_1n_2}T_2^{n_2}T_1^{n_1},$
where we have again used the commutativity $[T_1,T_2]=0$.

Let now $g(x)$ be a fixed function in $\Hil$, and let us define the
following functions: \be
g_{\underline{n}}^{(L)}(x):=T_1^{n_1}T_2^{n_2}\,g(x)=T_2^{n_2}T_1^{n_1}
\,g(x) =(-1)^{Ln_1n_2}U(\underline
n)\,g(x)=(-1)^{Ln_1n_2}D(\underline n)\,g(x).\label{45}\en We call
$\C^{(L)}=\{g_{\underline{n}}(x),\,\underline{n}\in{\Bbb{Z}}^2\}$
the set of these vectors, which is the standard set of coherent
states if $g(x)$ is just the vacuum of the operator $b$. These
functions, which quite often in the literature are written as
$e^{in_1\omega_0x}g(x-n_2x_0)$ for some positive $x_0$ and
$\omega_0$, are not mutually orthogonal for a generic $g(x)$. For
instance, they are not orthogonal for coherent states. Here we will
assume always that the functions in $\C^{(L)}$ are linearly
independent.

Our aim is to construct a family of vectors $\E^{(L)}$ which shares
with $\C^{(L)}$ most of its features and which, moreover, is made of
orthonormal vectors: we are specializing our original settings of
Section II to $N=2$. Of course, we are interested here in
considering a starting function $g(x)$ such that
$<g_{\underline{n}}^{(L)},g>\neq
\delta_{\underline{n},\underline{0}}$. Also, we would like to know
if for some choice of $g(x)\in\Lc^2(\Bbb{R})$ and $L\geq 1$ it is
possible that $C^{(L)}$ is a complete set in $\Lc^2(\Bbb{R})$. In
other words, if we define $\Hil_{g,L}:=\overline{\mbox{linear span
}\{g_{\underline{n}}^{(L)}(x),\,\underline{n}\in\Bbb{Z}^2\}}$, we
wonder whether a clever choice of $g$ and $L$ produces
$\Hil_{g,L}=\Lc^2(\Bbb{R})$.  These questions can be answered rather
easily by making use of the generalized $(k,q)$-representation.

\vspace{2mm}

We first remark that, introducing the generalized common eigenstates
of $T_1$ and $T_2$ as in the beginning of this section, we have
$T_1\Phi_{k,q}^{(A,a)}=e^{iqa}\Phi_{k,q}^{(A,a)}$ and
$T_2\Phi_{k,q}^{(A,a)}=e^{ikA}\Phi_{k,q}^{(A,a)}$. Therefore
$<g_{\underline{n}}^{(L)},\Phi_{k,q}^{(A,a)}>=e^{-in_1aq}\,e^{-ikn_2A}<g,\Phi_{k,q}^{(A,a)}>$
which, in turns, using the resolution of the identity in (\ref{43}),
implies that
 \be
I_{g,\underline{n}}^{(L)}=<g_{\underline{n}}^{(L)},g>=\int_0^{2\pi/A}\,dk\,
e^{-ikn_2A}\,\int_0^a\,dq\,e^{-iqn_1a}\,\left|g(k,q)\right|^2\label{46}\en
where we have introduced the $(k,q)$ representation of the function
$g(x)$ as $g(k,q)=<\Phi_{k,q}^{(A,a)},g>$. We will now see that, as
in \cite{bagtri}, the role of $L$ is crucial here.

If $L=1$ the set
$\Sc^{(L)}=\{e^{-ikn_2A}\,e^{-iqn_1a},\,(n_1,n_2)\in\Bbb{Z}\}$ is
complete in $\Lc^2(\Box^{(A)})$, so that the functions in $C^{(1)}$
can be mutually orthogonal if and only if $g(k,q)$ is a phase so
that $|g(k,q)|$ is constant almost everywhere (a.e.) in
$\Box^{(A)}$. This is the reason why the coherent states are not
mutually orthogonal: the $(k,q)$ representation of the vacuum of the
annihilation operator $b$  is not simply a phase!

If $L>1$ the conclusion changes. Let us consider, as an example,
$L=2$. In this case the set $\Sc^{(2)}$ is no longer complete in
$\Lc^2(\Box^{(A)})$ so that equation (\ref{46}) does not
automatically implies that $|g(k,q)|$ must be constant. On the
contrary, splitting the double integral on $\Box^{(A)}$ in two
contributions over
$(k,q)\in[0,2\pi/A[\times[0,a/2[=:\Box^{(A)}_{1/2}$ and
$(k,q)\in[0,2\pi/A[\times[a/2,a[$, we can write
$I_{g,\underline{n}}^{(2)}$ as
$$
I_{g,\underline{n}}^{(2)}=\int_0^{2\pi/A}\,dk\,
e^{-ikn_2A}\,\int_0^{a/2}\,dq\,e^{-iqn_1a}\,\left(\left|g(k,q)\right|^2+\left|g(k,q+a/2)
\right|^2\right),
$$
which is equal to $\delta_{\underline{n},\underline{0}}$ if and only
if $ \left|g(k,q)\right|^2+\left|g(k,q+a/2) \right|^2$ is constant
almost everywhere for $(k,q)\in \Box^{(A)}_{1/2}$ since $\Sc^{(2)}$
is complete in $\Lc^2(\Box^{(A)}_{1/2})$. This result is easily
generalized:

\begin{prop} The set $C^{(L)}$ generated by a given square-integrable
function $g(x)$ is made of orthonormal functions if and only if in
the $(k,q)$-representation $g$ satisfies the following equality
almost everywhere \be
\left|g(k,q)\right|^2+\left|g\left(k,q+\frac{a}{L}\right) \right|^2
+\left|g\left(k,q+\frac{2a}{L}\right)
\right|^2+\cdots+\left|g\left(k,q+\frac{(L-1)a}{L}\right)
\right|^2=\frac{LA}{2\pi a} \label{47}\en \end{prop}Of course, once
such a $g(k,q)$ has been found, then the related $g(x)$ can be
simply obtained as \be
g(x)=\int\int_{\Box^{(A)}}\,dk\,dq\,\Phi_{k,q}^{(A,a)}(x)\,g(k,q)
\label{48}\en Any $g(x)$ which cannot be written in this form cannot
produce an o.n. set $\C^{(L)}$. As a consequence, this is exactly
the kind of functions we will consider in the rest of this section,
since we are interested here in producing an orthonormal set
starting from a different  set $\C^{(L)}$ which is not orthogonal
from the very beginning.

\vspace{2mm}

The generalized $(k,q)$ representation can also be used to discuss
the problem of the completeness of $\C^{(L)}$ in $\Lc^2(\Bbb{R})$.
Before doing this, however, it may be worth observing that by
definition $\C^{(L)}$ is complete in $\Hil_{g,L}$ but, since in
general $\Hil_{g,L}$ is only a subset of $\Lc^2(\Bbb{R})$, we have
no information about the completeness of $\C^{(L)}$ on this larger
Hilbert space. In other words, if we find conditions for $\C^{(L)}$
to be complete in $\Lc^2(\Bbb{R})$ we also find conditions for
$\Hil_{g,L}$ to coincide with $\Lc^2(\Bbb{R})$.

To answer these questions we first recall that the set $C^{(L)}$ is
complete in $\Lc^2(\Bbb{R})$ if, given a square integrable function
$h(x)$ which is orthogonal to $g_{\underline{n}}^{(L)}(x)$ for all
$\underline{n}\in\Bbb{Z}^2$, then $h(x)=0$ a.e. in $\Bbb{R}$. Using
the properties of the functions $\Phi_{k,q}^{(A,a)}$ we can write
$$
<g_{\underline{n}}^{(L)},h>=\int_0^{2\pi/A}\,dk\,
e^{-ikn_2A}\,\int_0^{a}\,dq\,e^{-iqn_1a}\,\overline{g(k,q)}\,h(k,q),
$$
where, as usual, $g(k,q)=<\Phi_{k,q}^{(A,a)},g>$ and
$h(k,q)=<\Phi_{k,q}^{(A,a)},h>$. Again, to find conditions for these
scalar products to be zero $\forall\,\underline{n}\in\Bbb{Z}^2$, it
is convenient to consider separately the two cases: $L=1$ and $L>1$.

If $L=1$, using as before the completeness of the set $\Sc^{(1)}$ in
$\Lc^2(\Box^{(A)})$, we deduce that $<g_{\underline{n}}^{(L)},h>=0$
$\forall\,\underline{n}\in\Bbb{Z}^2$ if and only if
$\overline{g(k,q)}\,h(k,q)=0$ a.e. in $\Box^{(A)}$. Therefore, if
$g(k,q)$ is zero at most on a set of zero measure, this implies that
$h(k,q)=0$ a.e. in $\Box^{(A)}$ and, as a consequence, that $h(x)=0$
a.e. in $\Bbb{R}$.

\vspace{2mm}

\noindent{\bf Remark:} the vacuum of $b$, which generates the set of
coherent states, satisfies this condition, \cite{bgz}, and therefore
the related set $\C^{(1)}$ is complete in $\Lc^2(\Bbb{R})$. Examples
of functions which {\bf do not} satisfy this condition can be easily
constructed. Consider, for instance, a vector $\hat g$ which in the
$(k,q)$-representation is equal to 1 for $(k,q)\in\Box^{(A)}_{1/2}$
and zero otherwise. In the $x$-representation this function looks
like $\hat g(x)=\sqrt{\frac{2\pi}{A}}\,\chi_{[0,a/2[}(x)$, where
$\chi_{I}(x)$ is, as usual,  the characteristic function of the set
$I$. For what we have just shown this function does not generate a
complete set in $\Lc^2(\Bbb{R})$ since non zero functions $h(k,q)$
for which $\overline{g(k,q)}\,h(k,q)=0$ a.e. in $\Box^{(A)}$ can be
easily found.

\vspace{2mm}

Let us now consider what happens for $L>1$ and, to be concrete, let
us fix $L=2$. Splitting the integral as we have done before, we
deduce that $<g_{\underline{n}}^{(L)},h>=0$ for all
$\underline{n}\in\Bbb{Z}^2$ if and only if  is zero  the following
combination $
\overline{g(k,q)}\,h(k,q)+\overline{g\left(k,q+\frac{a}{2}\right)}\,h\left(k,q+\frac{a}{2}\right).
$ In other words, $<g_{\underline{n}}^{(L)},h>=0$ for all
$\underline{n}\in\Bbb{Z}^2$ if and only if \be
\overline{g(k,q)}\,h(k,q)+\overline{g\left(k,q+\frac{a}{2}\right)}\,h\left(k,q+\frac{a}{2}\right)=0\quad
\mbox{ a.e. in } \Box^{(A)}_{1/2}\label{49}\en and, as a
consequence, $\C^{(L)}$ is complete in $\Lc^2(\Bbb{R})$ for some
given $g(k,q)$ if and only if the only solution of equation
(\ref{49}) is $h(k,q)=0$ a.e. in $\Box^{(A)}_{1/2}$. Of course, any
function $g(k,q)$ which is zero on a set $\D\subset\Box^{(A)}_{1/2}$
of non zero measure  cannot produce a complete set, because equation
(\ref{49}) would have a non trivial solution: it is enough to choose
a function $h(k,q)$ which is zero only outside $\D$! Hence, in order
to obtain a complete set, it is surely necessary to assume  that
$g(k,q)$ is zero at most on a set of zero measure. However, even
under this assumption it is easy to check that other non trivial
solutions of this equation do exist, and therefore $\C^{(L)}$ cannot
be complete in $\Lc^2(\Bbb{R})$. Let indeed $s_o(x)$ be a fixed
function in $\Lc^2(\Bbb{R})$ and let
$\mu(k,q):=e^{iqa/2}\,\sum_{n\in\Bbb{Z}}\,e^{inkA}\,s_o(q+na/2)$.
$\mu(k,q)$ belongs to $\Lc^2(\Box^{(A)}_{1/2})$ since
$\int\int_{\Box^{(A)}_{1/2}}\,|\mu(k,q)|^2\,dk\,dq=\frac{2\pi}{A}\,\int_{\Bbb{R}}\,|s_o(x)|^2\,dx$,
which is finite because $s_o(x)\in \Lc^2(\Bbb{R})$. Using now the
 boundary conditions $g(k,q+a)=e^{ikA}\,g(k,q)$ and
$\mu(k,q+a/2)=-e^{ikA}\,\mu(k,q)$, it is easy to check that
$h(k,q):=\overline{g(k,q+a/2)}\,\mu(k,q)$ satisfies equation
(\ref{49}) and is different from zero a.e. if $s_o(x)$ is chosen,
e.g.,  positive, since in this case $g(k,q)\neq0$ a.e. in
$\Box^{(A)}_{1/2}$.

Let us now summarize the above results: {\em the set $\C^{(L)}$
consists of o.n. functions if and only if $g(k,q)$ satisfies
equation (\ref{47}). Moreover, if $g(k,q)\neq0$ a.e. in
$\Box^{(A)}$, then $\C^{(1)}$ is complete in $\Lc^2(\Bbb{R})$.
Finally, if $L>1$, there is no choice of $g(k,q)$ for which the set
$\C^{(L)}$ is complete in $\Lc^2(\Bbb{R})$. Therefore, for any given
$g$ and for all $L\geq 2$ we have
$\Hil_{g,L}\subset\Lc^2(\Bbb{R})$.}

This result reflects, in a certain sense, what happens for frames of
translates  which are surely not complete in all of
$\Lc^2(\Bbb{R})$, as we have discussed in the previous section.

\subsection{Back to the orthonormalization problem}

We are now ready to apply our {\em orthonormalization procedure} to
the set $g_{\underline{n}}(x)$ in (\ref{45}), with the only
requirement that $\C^{(L)}$ must consists of functions which are not
orthogonal from the very beginning and  linearly independent. As in
\cite{bagtri} and in Section II we define \be
\Psi_{\underline{n}}^{(L)}(x):=\sum_{\underline{k}\in{\Bbb{Z}}^2}\,c_{\underline{k}}^{(L)}\,
g_{\underline{k}+\underline{n}}^{(L)}(x), \label{410} \en which
implies that
$\Psi_{\underline{0}}^{(L)}(x):=\sum_{\underline{k}\in{\Bbb{Z}}^2}\,c_{\underline{k}}^{(L)}\,
g_{\underline{k}}^{(L)}(x)$ and, because of the commutativity of
$T_1$ and $T_2$, that \be
\Psi_{\underline{n}}^{(L)}(x)=T_1^{n_1}\,T_2^{n_2}\Psi_{\underline{0}}^{(L)}(x).
\label{411m}\en Therefore the new set constructed in this way,
$\E^{(L)}:=\{\Psi_{\underline{n}}^{(L)}(x),\,\underline{n}\in{\Bbb{Z}}^2\}$,
is invariant under the action of $T_1$ and $T_2$, exactly as the set
$\C^{(L)}$, independently of the choice of the coefficients of the
expansion $c_{\underline{k}}^{(L)}$.  These coefficients
$c_{\underline{s}}^{(L)}$  are, as usual, fixed (non uniquely) by
requiring that the vectors in the set $\E^{(L)}$ are orthonormal,
$<\Psi_{\underline{n}}^{(L)},\Psi_{\underline{s}}^{(L)}>=\delta_{\underline{n},
\underline{s} }$, or, equivalently, that
$<\Psi_{\underline{n}}^{(L)},\Psi_{\underline{0}}^{(L)}>=\delta_{\underline{n},
\underline{0} }$.  Indeed, if they exist, the coefficients can be
found as before: \be
c_{\underline{k}}^{(L)}=\frac{1}{(2\pi)^2}\,\int_0^{2\pi}\int_0^{2\pi}\,\frac{e^{-i\underline{P}
\cdot\underline{k}}}{\sqrt{F_L(\underline{P})}}\,d\underline{P},\label{411}\en
where
$F_L(\underline{P})=\sum_{\underline{m}\in{\Bbb{Z}}^2}\,I_{\underline{m}}^{(L)}\,e^{i\underline{P}\cdot\underline{m}}$.
As in \cite{bagtri}, the behavior of the coefficients in (\ref{411})
is directly related to the nature of the convergence and to the
zeros of $F_L(\underline{P})$, which, in turns, are related to the
coefficients of the overlap between $g_{\underline{m}}^{(L)}(x)$ and
$g(x)$, $I_{\underline{m}}^{(L)}=<g_{\underline{m}}^{(L)},g>$. In
particular the following results immediately follows from basic
facts in the Fourier series analysis and from the results in
\cite{bagtri}:

\begin{enumerate}

\item if $I_{\underline{m}}^{(L)}\in l^1(\Bbb{Z}^2)$ then
$F_L(\underline{P})$ is continuous, non negative,
$(2\pi,2\pi)$-periodic and bounded;

\item if $I_{\underline{m}}^{(L)}\in s(\Bbb{Z}^2)$ then
$F_L(\underline{P})$ is a $C^\infty$ non negative function;

\item if $I_{\underline{m}}^{(L)}$ are such that $\sum_{\underline{m}
\in\Bbb{Z}^2}\,\left|I_{\underline{m}}^{(L)}\right|<1$ then
$F_L(\underline{P})\neq 0$ for all
$\underline{P}\in[0,2\pi[\times[0,2\pi[$ and, as a consequence,
$\{c_{\underline{m}}^{(L)}\}\in s(\Bbb{Z}^2)$;

\item if $I_{\underline{m}}^{(L)}\in s(\Bbb{Z}^2)$ then
the sequence $\Psi_{\underline{n},\,N}^{(L)}:=\sum_{\underline{k}:
\|\underline{k}\|\leq
N}\,c_{\underline{k}}^{(L)}g_{\underline{k}}^{(L)}$ converges in the
$\|.\|_2$-norm so that its limit, $\Psi_{\underline{n}}^{(L)}$,
belongs to $\Hil_{g,L}$.

\end{enumerate}

If we introduce the operator $X_L=\sum_{\underline{m}
\in\Bbb{Z}^2}\,c_{\underline{m}}^{(L)}\,T_1^{m_1}\,T_2^{m_2}$ then
$\Psi_{\underline{n}}^{(L)}=X_L\,g_{\underline{n}}^{(L)}$. It is
also possible to extend the following proposition, originally given
in \cite{bagtri}:

\begin{prop}
Suppose that $\{c_{\underline{m}}^{(L)}\}\in l^1(\Bbb{Z}^2)$ and
that $F_L(\underline{P})\neq 0$ for all
$\underline{P}\in[0,2\pi[\times[0,2\pi[$. Then $\E^{(L)}$ is
complete in $\Hil_{g,L}$ if and only if $X_L$ admits a bounded
inverse.
\end{prop}
The proof is very close to that given in \cite{bagtri} and will not
be repeated here.

It is rather interesting to observe that the same sum rule which was
proved, under the assumptions of this Proposition, in \cite{bagtri},
that is
$\sum_{\underline{k}\,\in{\Bbb{Z}}^2}\,\overline{\alpha_{\underline{k}}^{(L)}}\,c_{\underline{k}}^{(L)}=1$,
also holds true in the present settings. Here the
$\alpha_{\underline{k}}^{(L)}$'s are the coefficients of the
expansion of the non-orthogonal functions $g_{\underline{n}}^{(L)}$
in terms of the orthogonal ones, $\Psi_{\underline{n}}^{(L)}$:
$g_{\underline{n}}^{(L)}(x)=\sum_{\underline{m}
\in\Bbb{Z}^2}\,\alpha_{\underline{m}}^{(L)}\,\Psi_{\underline{n}+\underline{m}}^{(L)}(x)$.
Explicitly we find
$\alpha_{\underline{m}}^{(L)}=\frac{1}{(2\pi)^2}\,\int_0^{2\pi}\int_0^{2\pi}e^{-i\underline
P\cdot\underline m}\,\sqrt{F_L(\underline P)}$. Again, we don't give
here the proof of this result which is based on the Poisson
summation formula since it is very close to that given in
\cite{bagtri}. We just want to stress that for each starting
function $g(x)$ we get a different sum rule.

The vectors of the set $\E^{(L)}$ have the following properties: \be
\sum_{\underline{n}\,\in{\Bbb{Z}}^2}\,|\Psi_{\underline{n}}^{(L)}><\Psi_{\underline{n}}^{(L)}|=
\sum_{\underline{n}\,\in{\Bbb{Z}}^2}\,|g_{\underline{n}}^{(L)}><g_{\underline{n}}^{(L)}|=\1_{g,L},\label{412}
\en where $\1_{g,L}$ is the identity operator in $\Hil_{g,L}$,
which, for what we have proven at the beginning of this section, can
be the identity in $\Lc^2(\Bbb{R})$ only if $L=1$ and $g$ is taken
conveniently.

Also, if by chance $g_{\underline{n}}^{(L)}(x)$ is an eigenstate of
some operator $\hat k_L$ (as it happens for  coherent states), then
$\Psi_{\underline{n}}^{(L)}(x)$ is an eigenstate of the operator
$\hat K_L=X_L\,\hat k_L\,X_L^{-1}$, with the same eigenvalue.

Here as in \cite{bagtri} the role of $L$ is very important and in
particular we are not sure a priori that $X_L^{-1}$ exists for all
values of $L$. For instance, if $L=1$ it is possible to adapt the
same {\em reductio ad absurdum} argument given in \cite{bagtri} to
conclude that our orthogonalization procedure  cannot work. More in
detail, suppose that we have constructed the o.n. set $\E^{(1)}$
which is a basis of $\Lc^2(\Bbb{R})$. Then we conclude, as in
\cite{bagtri}, that the original set $\C^{(1)}$  is an o.n. basis of
$\Lc^2(\Bbb{R})$ as well, which is false.

\vspace{2mm}

Before considering some examples a final remark is in order: for
Gabor frames it is widely discussed in the literature, see
\cite{dau} for instance and references therein, that 3 {\em
different regions} appear in their analysis and they give rise to
different mathematics. More in details: the set $\{g_{\underline
n}(x)=e^{ian_1x}g(x+an_2),\,n_1,n_2\in\Bbb{Z}\}$ can be an
orthonormal basis only if $a^2=2\pi$ (but with poor localization
properties). It can be a frame for all of $\Hil$ if $a^2<2\pi$ (and
it may have good localization properties). Finally, it cannot be a
frame for all of $\Hil$ if $a^2>2\pi$. Since our strategy works for
$a^2=2\pi L$, $L=1,2,3,\ldots$, this automatically excludes the case
$a^2<2\pi$. This constraint, in our approach, is due to the
important requirement that $[T_1,T_2]=0$ which we hope to weaken in
a close future. However, as we have widely discussed here and in
\cite{bagtri}, already for $a^2\geq 2\pi$ we find a very rich
mathematical setting, which produces many interesting results.

\section{Examples and conclusions}

The first example of this construction produces the {\em classical}
coherent states. We don't discuss this example here since few
results have already been given in Section II and a complete
analysis can be found in \cite{bagtri}. In this paper  we
concentrate our attention on other examples whose computations can
be carried out almost completely analytically. Other examples can be
discussed in a totally analogous way.

{\bf Example 1.}

Let us consider the following normalized compactly supported
function,
$$
g(x)=\left\{
\begin{array}{ll} \frac{1}{\sqrt{3a/2}},\quad |x|\leq \frac{3a}{4}\\
0,\hspace{1.1cm}\mbox{ otherwise.}\end{array} \right.
$$
It is clear that the different $g_{\underline
n}(x)=e^{ian_1x}g(x+an_2)$ are not all automatically orthogonal,
even if $<g_{\underline n},g_{\bf 0}>$ is surely zero if $|n_2|\geq
2$ because, in this case, the supports of the two functions have
empty intersection. It is easy to check that, choosing $L=4$ so that
$a^2=8\pi$ to simplify the computations, $F_L(\underline
P)=1+\frac{2}{3}\,\cos(P_2)=:F_L(P_2)$ which is independent of
$P_1$. As a consequence, the coefficients $c_{\underline{k}}^{(L)}$
in (\ref{411}) can be written as
$c_{\underline{k}}^{(L)}=\delta_{k_1,0}\,\hat c_{k_2}$, where
$$
\hat
c_{k_2}=\frac{1}{2\pi}\int_0^{2\pi}\frac{e^{-iP_2k_2}}{\sqrt{F_L(P_2)}}\,dP_2.
$$
These coefficients can be easily computed: $\hat c_0=1.11308 $,
$\hat c_{\pm 1}=-0.216769$, $\hat c_{\pm 2}=0.0625106$ and so on.
Needless to say, because of the analytic properties of $F_L(P_2)$
which is never zero in $[9,2\pi[$, they tend to zero faster than any
inverse power of $k_2$ so that all the series which appear along our
computations are surely convergent (in the strongest topology).

From formula (\ref{410}) we find that $\Psi(x)=\sum_{k}\hat
c_k\,g(x+ka)$. We plot in the following figure
$\Psi_N(x)=\sum_{k=-N}^N\,\hat c_k\,g(x+ka)$ for different values of
$N$.

\begin{center}
\mbox{\includegraphics[height=3.2cm, width=4.5cm]
{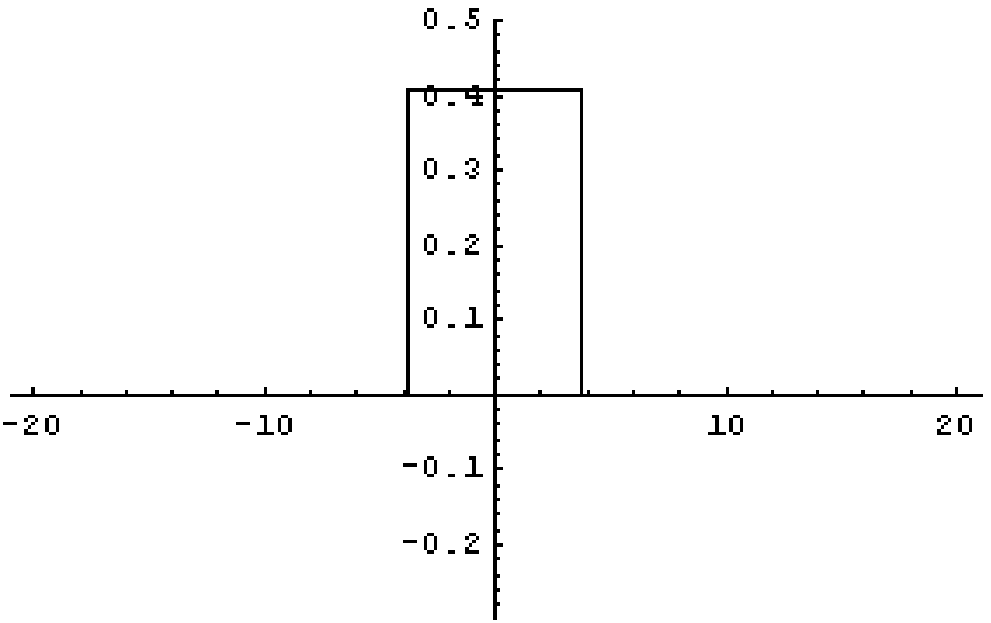}}\hspace{6mm}
\mbox{\includegraphics[height=3.2cm, width=4.5cm]
{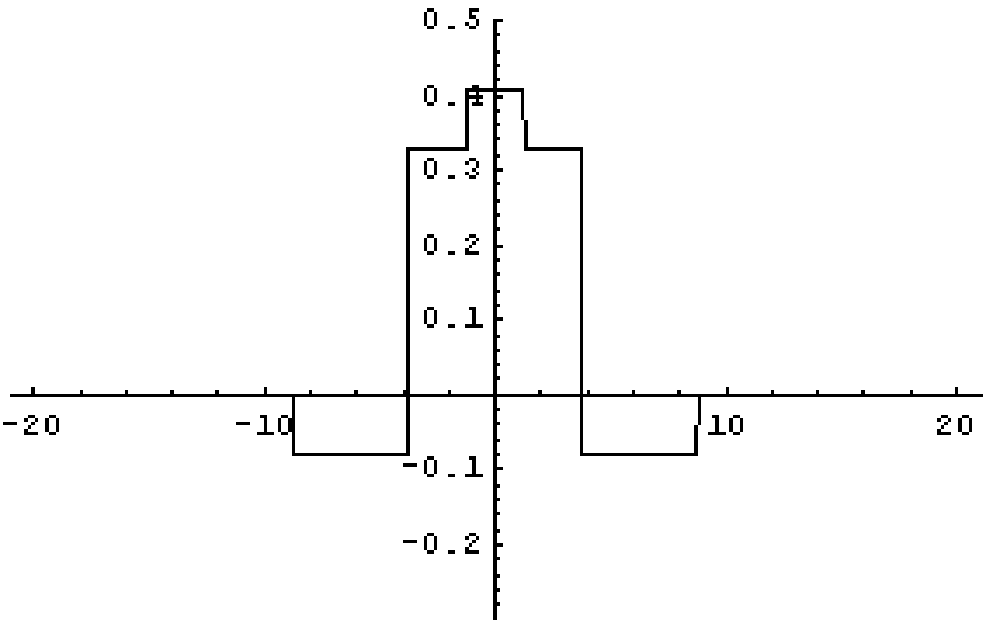}}\hspace{6mm}
\mbox{\includegraphics[height=3.2cm, width=4.5cm]
{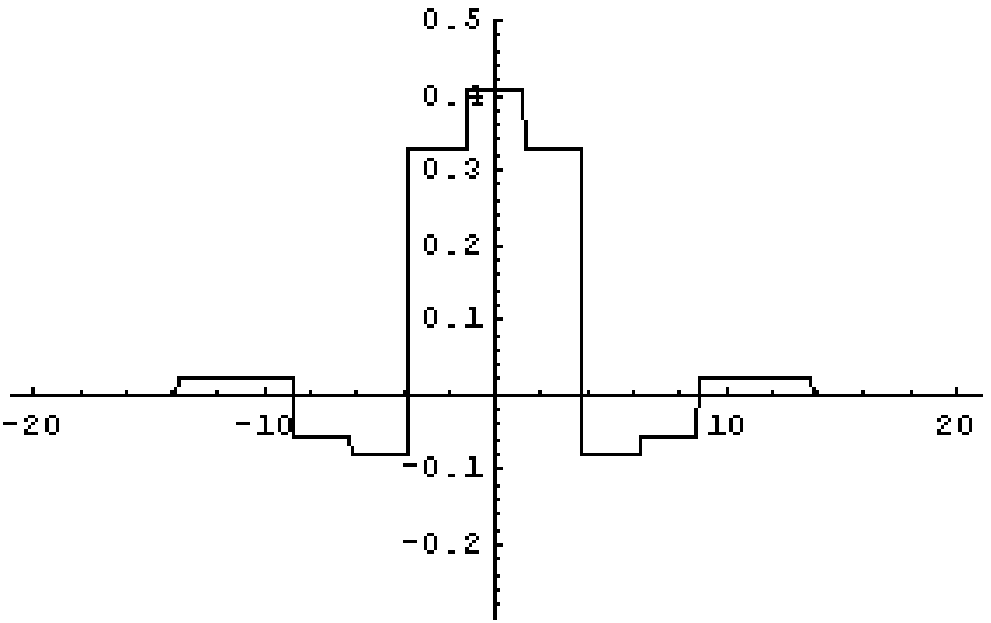}} \hspace{6mm}
\mbox{\includegraphics[height=3.2cm, width=4.5cm]
{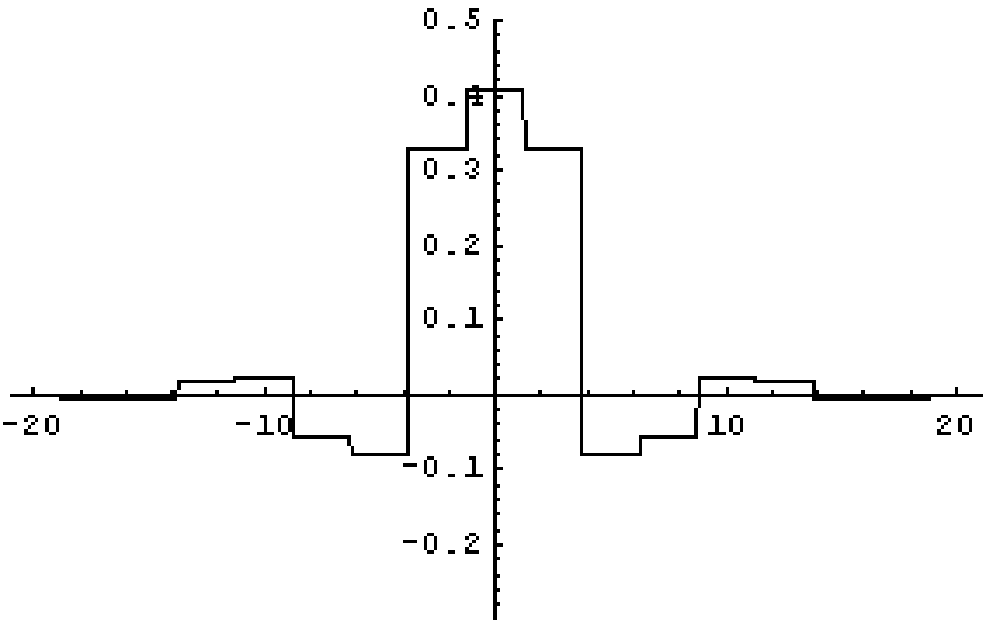}}\hspace{6mm}
\mbox{\includegraphics[height=3.2cm, width=4.5cm]
{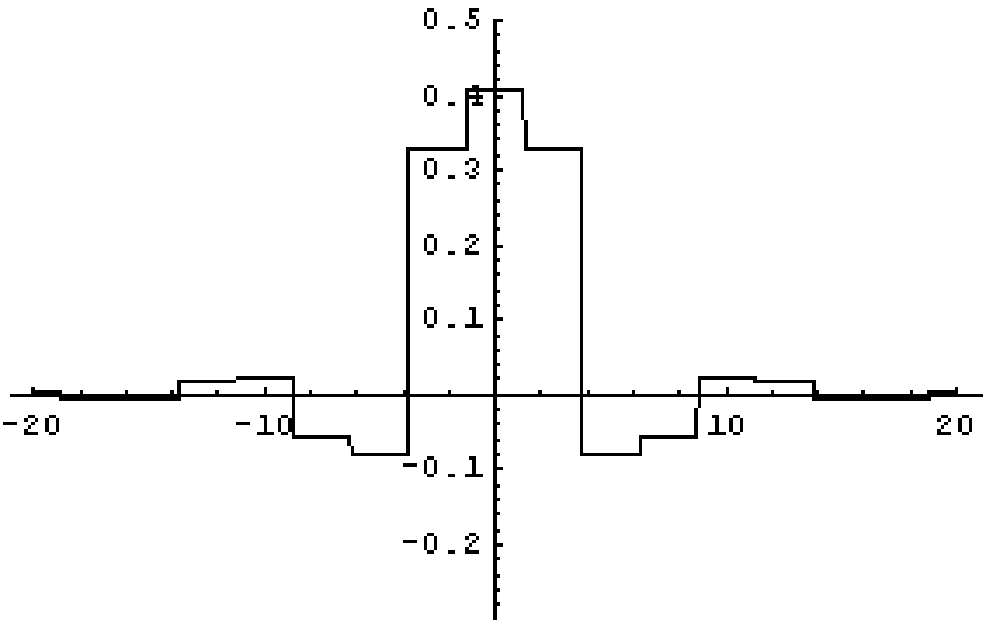}} \hspace{6mm}
\hfill\\
\begin{figure}[h]
\caption{\label{fig1} $\Psi(x)$ for $N=0$, $N=1$, $N=2$ (first row)
and $N=3$, $N=4$ (second row) }
\end{figure}
\end{center}

It is just a technicality to check that the overlap between
$\Psi(x)$ and $T_1^{n_1}T_2^{n_2}\Psi(x)$ can be written in terms of
the coefficients $\hat c_k$ as follows: \be <\Psi_{\underline
n}^{(L)},\Psi>=\sum_{l\in\Bbb{Z}}\left[\hat c_l(\hat c_{l+n_2}+(\hat
c_{l+n_2+1}+\hat c_{l+n_2-1})/3)\right]\,\delta_{n_1,0}
\label{413}\en and we find that, replacing $\sum_{l\in\Bbb{Z}}$ with
$\sum_{l=-N}^N$ in (\ref{413}), for $N=3$ we have
$\|\Psi\|^2=1.00001$ while the modulus of the overlap between two
different wave-functions does not exceed  $0.00605858$. Therefore we
conclude that this is already a very good approximation, which can
however be improved if we take $N=4$: in this case we find
$\|\Psi\|^2=1$ and the modulus of the overlap between two different
wave-functions is always less than $0.00208293$.

The computation of $X_L$ directly follows from its definition:
$X_L=\sum_{n_2\in\Bbb{Z}}\hat c_{n_2}^{(L)}\,T_2^{n_2}$. Its
inverse, $X_L^{-1}$, can be computed in complete analogy: since
$g_{\underline{n}}^{(L)}(x)=\sum_{\underline{m}
\in\Bbb{Z}^2}\,\alpha_{\underline{m}}^{(L)}\,\Psi_{\underline{n}+\underline{m}}^{(L)}(x)=X_L^{-1}
\Psi_{\underline{n}}^{(L)}(x)$, we get
$X_L^{-1}=\sum_{n_2\in\Bbb{Z}}\hat \alpha_{n_2}^{(L)}\,T_2^{n_2}$,
where $\hat
\alpha_{n_2}^{(L)}=\frac{1}{2\pi}\int_0^{2\pi}dP\,e^{-iPn_2}\sqrt{F_L(P)}$.
We find, for instance, $\hat \alpha_{0}^{(L=4)}=0.96857$, $\hat
\alpha_{\pm 1}^{(L=4)}=0.175095$, $\hat \alpha_{\pm
2}^{(L=4)}=-0.016400$, and so on.

\vspace{3mm}

{\bf Example 2.}

Let us consider now the following function:
$$
g(x)=\left\{
\begin{array}{ll} N_b\,\exp\left(\frac{1}{x^2-b^2}\right),\quad |x|\leq b\\
0,\hspace{2.7cm}\mbox{ otherwise,}\end{array} \right.
$$
where $b=\frac{3a}{4}$, $a^2=2\pi L$, has been introduced only to
simplify the notation. $N_b$ is a normalization constant which
depends obviously on $b$ and, as a consequence, on $L$. Once again
we consider a function with compact support just to simplify the
computations, since with our choice $<g_{\underline n},g>$ is zero
if $|n_2|\geq 2$. Hence, the function $F_L(\underline P)$ can be
written as $F_L(\underline P)=f_0(P_1)+2f_1(P_1)\,\cos(p_2)$, where
$f_0(P_1)=\sum_{n_1\in\Bbb{Z}}\,I_{n_1,0}^{(L)}\,e^{iP_1n_1}$ and
$f_1(P_1)=\sum_{n_1\in\Bbb{Z}}\,I_{n_1,1}^{(L)}\,e^{iP_1n_1}$. The
simplest results are obtained when both these functions can be {\em
reasonably well} replaced by their constant main contributions, that
is those contributions coming from $n_1=0$. Indeed, if this can be
done, then we get $F_L(\underline
P)\simeq1+2\,I_{(0,1)}^{(L)}\,\cos(P_2)=:F_L(P_2)$ which does not
depend on $P_1$, so that, once again,
$c_{\underline{k}}^{(L)}=\delta_{k_1,0}\,\hat c_{k_2}^{(L)}$, where
$ \hat
c_{k_2}^{(L)}=\frac{1}{2\pi}\int_0^{2\pi}\frac{e^{-iP_2k_2}}{\sqrt{F_L(P_2)}}\,dP_2
$ and $\Psi^{(L)}(x)=\sum_{k}\hat c_k^{(L)}\,g^{(L)}(x+ka)$. Of
course, replacing $f_0(P_1)$ and $f_1(P_1)$ respectively with their
first contributions $I_{0,0}^{(L)}=1$ and $I_{0,1}^{(L)}$ is
possible only if we have some control on what we are neglecting.
However, if $L=1$, it is easy to check that $I_{(0,1)}^{(L)}$ is
smaller than that part of $f_2(P_2)$ we are neglecting, $\delta
f_2(P_2)$, so that replacing $F_L(\underline P)$ with $F_L(P_2)$ is
a dangerous operation! This will be reflected on the fact that, as
we will see, the orthonormalization procedure does not work in this
case: we get a new family of functions which are not all mutually
orthogonal as we would like. On the other hand, if we consider
$L=2$, then $|\delta f_2(P_2)|<|I_{0,1}^{(L)}|$ but they are of the
same order of magnitude: we have some chance that the procedure
works but we cannot be sure at this stage. For $L\geq 3$, finally,
$|\delta f_2(P_2)|\ll|I_{0,1}^{(L)}|$ so that we expect to get a
{\em good} orthonormal set for $\Hil_{g,L}$.

These are exactly the conclusions that we obtain at the end of the
computations: if $L=1$ the  function $\Psi^{(1)}(x)=\sum_{k}\hat
c_k^{(1)}\,g^{(1)}(x+ka)$ is such that $\|\Psi^{(1)}\|^2\simeq
0.96$, while, for instance, $<\Psi^{(1)},\Psi_{(0,1)}^{(1)}>\simeq
0.1$. We see that the related functions do not produce  an
orthogonal set.

Let us see what happens if $L=2$. In this case we find
$\|\Psi^{(2)}\|^2\simeq 0.99996$ and
$<\Psi^{(2)},\Psi_{(0,1)}^{(2)}>\simeq 0.0001$, which are quite
promising. However, we also find that
$<\Psi^{(2)},\Psi_{(1,0)}^{(2)}>\simeq 0.037$ which shows that for
$L=2$ the approximated $\Psi_{\underline n}^{(2)}$'s constructed as
shown above are orthogonal up to corrections which are small but not
too much. However, as expected, this {\em out of orthogonality}
parameter gets smaller and smaller when $L$ increases: already for
$L=3$ we find $<\Psi^{(3)},\Psi_{(1,0)}^{(3)}>\simeq -0.012$, and
all the other scalar products are much smaller. The norm of
$\Psi^{(3)}$ is essentially 1. In the following figure we plots the
different approximations of $\Psi^{(L)}(x)$ for $L=3$.

\begin{center}
\mbox{\includegraphics[height=3.2cm, width=4.5cm]
{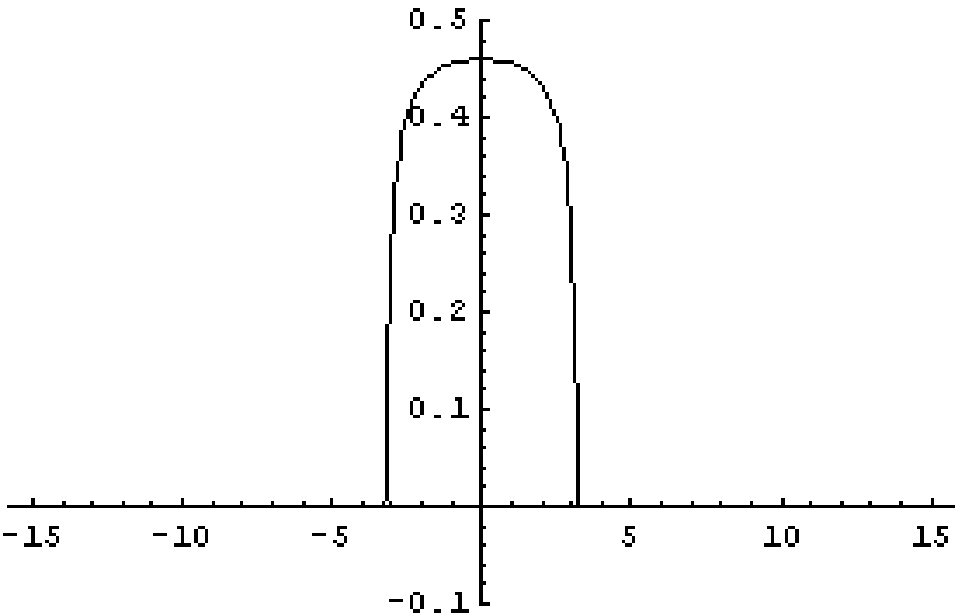}}\hspace{6mm}
\mbox{\includegraphics[height=3.2cm, width=4.5cm]
{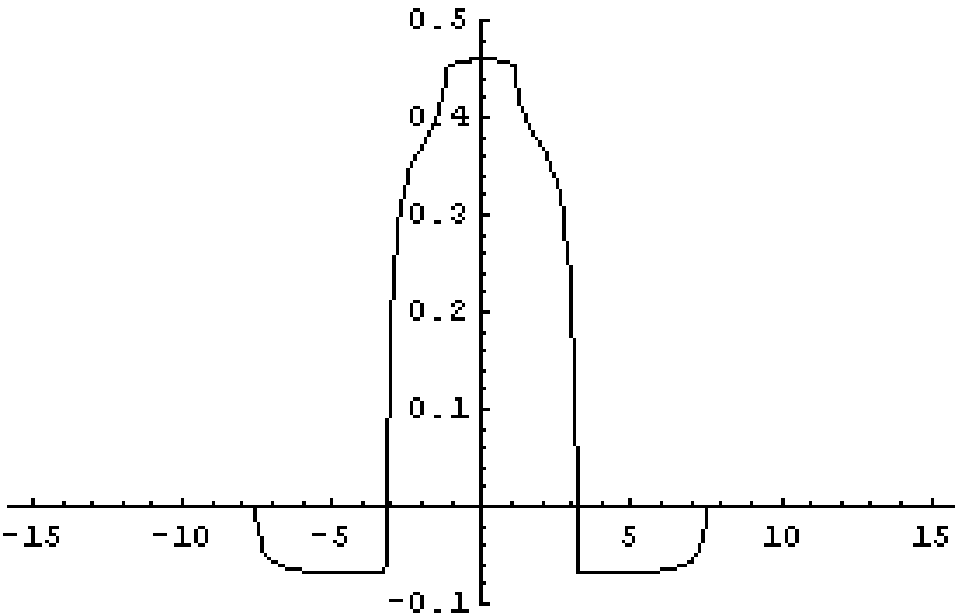}}\hspace{6mm}
\mbox{\includegraphics[height=3.2cm, width=4.5cm]
{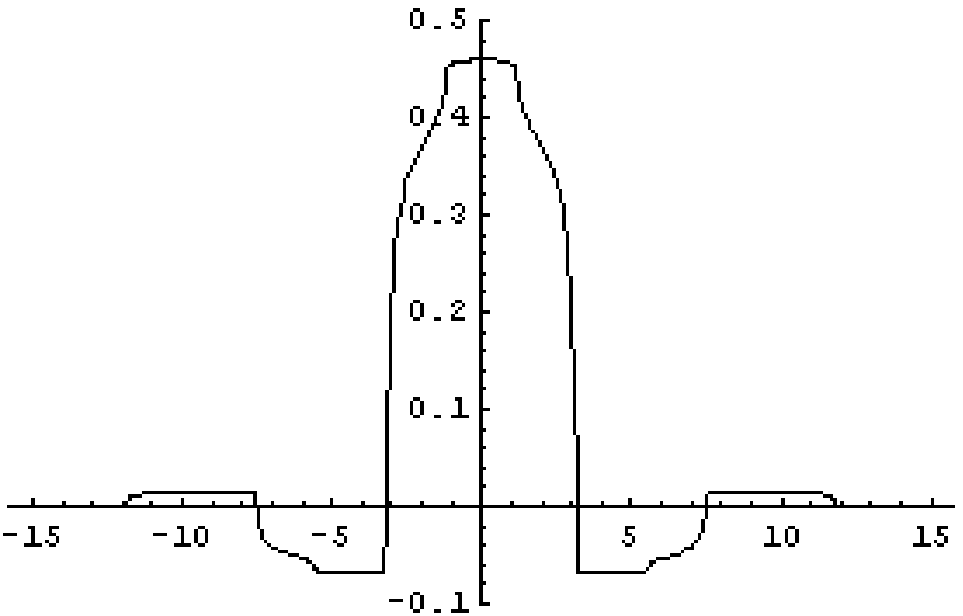}} \hspace{6mm}
\begin{figure}[h]
\caption{\label{fig2} $\Psi(x)$ for $N=0$, $N=1$, $N=2$  }
\end{figure}
\end{center}

\vspace{3mm}

{\bf Example 3.}

Let us consider now the following function:
$$
g(x)=\left\{
\begin{array}{ll} \frac{1}{\sqrt{a}}\,\cos\left(\frac{\pi x}{2a}\right),\quad |x|\leq a\\
0,\hspace{2.2cm}\mbox{ otherwise.}\end{array} \right.
$$
Also in this example we consider a function with compact support
since it allows to perform almost all the computations analytically.
As before $<g_{\underline n},g>$ is  zero if $|n_2|\geq 2$, and this
makes the computation of $F_L(\underline P)$ simple. Indeed, if we
introduce the function
$f(P_1)=\sum_{m_1\in\Bbb{Z}}\frac{e^{iP_1m_1}}{\pi-4\pi L^2 m_1^2}$
then we get $F_L(\underline P)=1+2\,f(P_1)\cos(P_2)$. Of course, due
to this formula, $c_{\underline n}^{(L)}$ is not the product of
$\delta_{n_1,0}$ and a coefficient which only depends on $n_2$.
Nevertheless, this is exactly what happens if we consider the first
approximation of $f(P_1)\simeq \frac{1}{\pi}$, in analogy with what
we have done in the previous example. Again, this replacement is
 justified especially for values of $L$ larger than 2 and  we find
$c_{\underline{k}}^{(L)}=\delta_{k_1,0}\,\hat c_{k_2}$, where $ \hat
c_{k_2}=\frac{1}{2\pi}\int_0^{2\pi}\frac{e^{-iP_2k_2}}{\sqrt{F_L(P_2)}}\,dP_2
$, $F_L(P_2)\simeq1+\frac{2}{\pi}\cos(P_2)$, and
$\Psi(x)=\sum_{k}\hat c_k\,g(x+ka)$.

The coefficients can again be easily computed: $\hat c_0=1.0997 $,
$\hat c_{\pm 1}=-0.20105$, $\hat c_{\pm 2}=0.0545131$ and so on. Of
course, because of the analytic properties of $F_L(P_2)$, they tend
to zero faster than any inverse power of $k_2$ so that even in this
example all the series which appear along our computations are
surely convergent.

We plot in the following figure
$\Psi_N(x)=\sum_{k=-N}^N\,c_k\,g(x+ka)$ for different values of $N$
and for $L=1$. Not many differences appear in the shapes of the
functions for larger values of $L$. The only major difference is a
bigger support.

\begin{center}
\mbox{\includegraphics[height=3.2cm, width=4.5cm]
{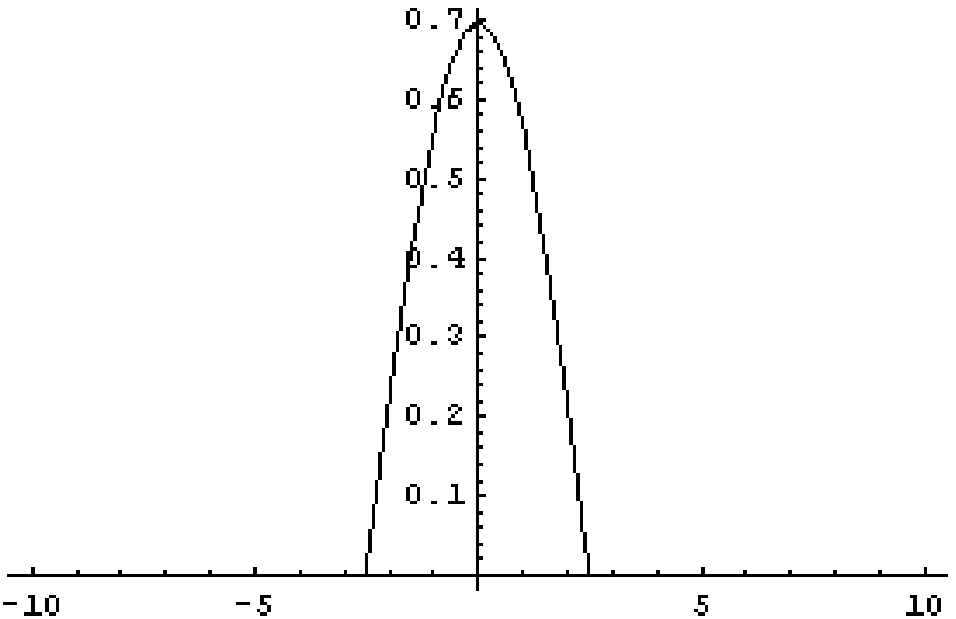}}\hspace{6mm}
\mbox{\includegraphics[height=3.2cm, width=4.5cm]
{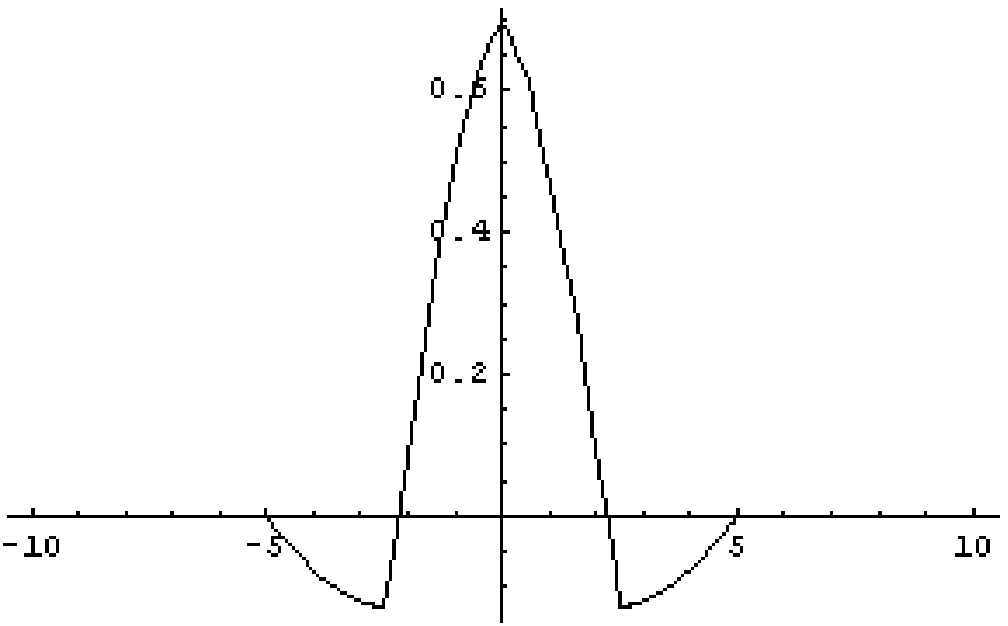}}\hspace{6mm}
\mbox{\includegraphics[height=3.2cm, width=4.5cm]
{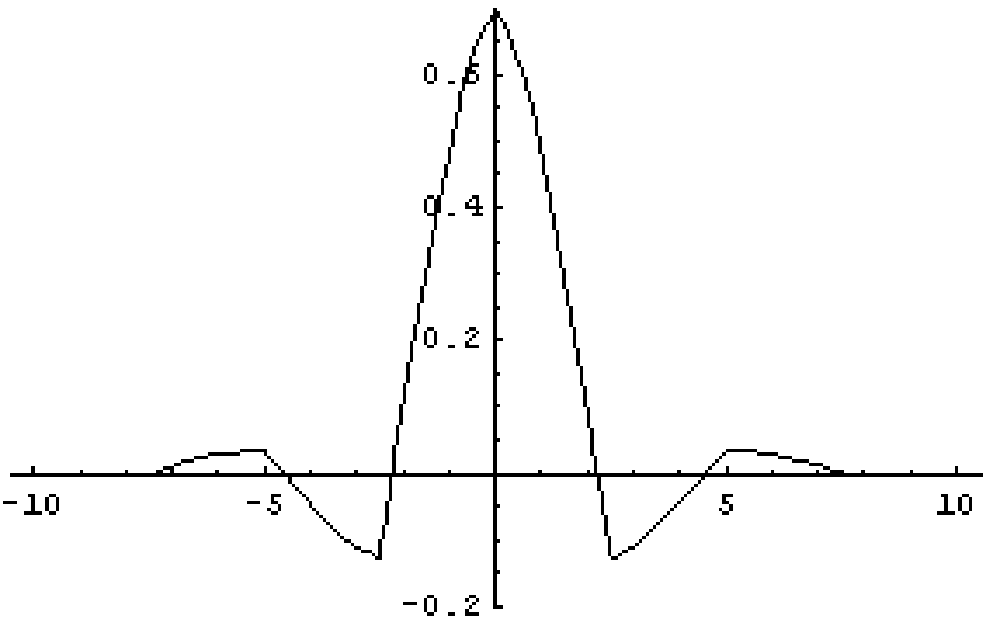}} \hspace{6mm}
\mbox{\includegraphics[height=3.2cm, width=4.5cm]
{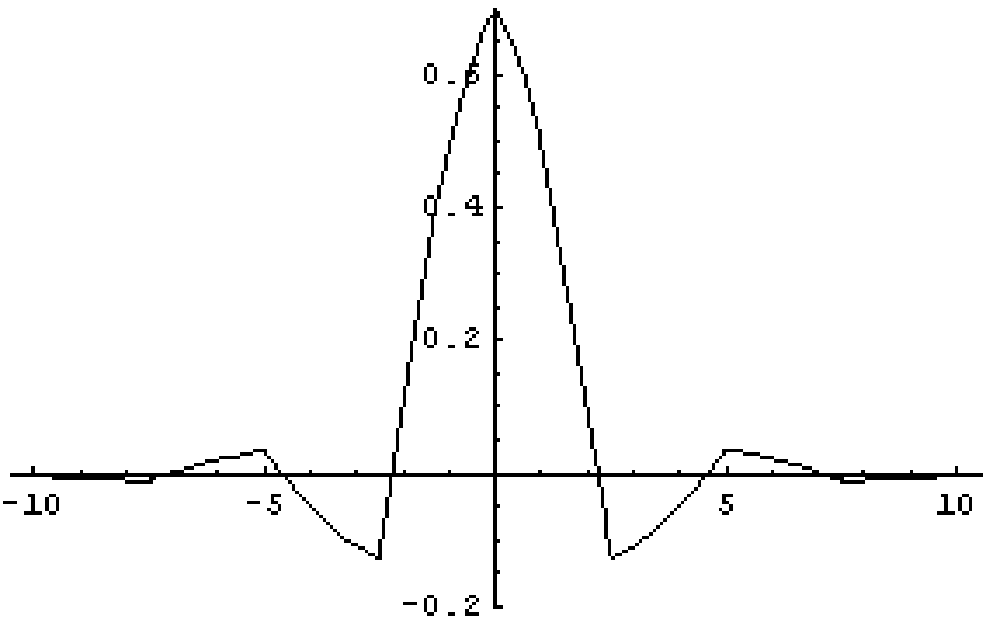}}\hspace{6mm}
\mbox{\includegraphics[height=3.2cm, width=4.5cm]
{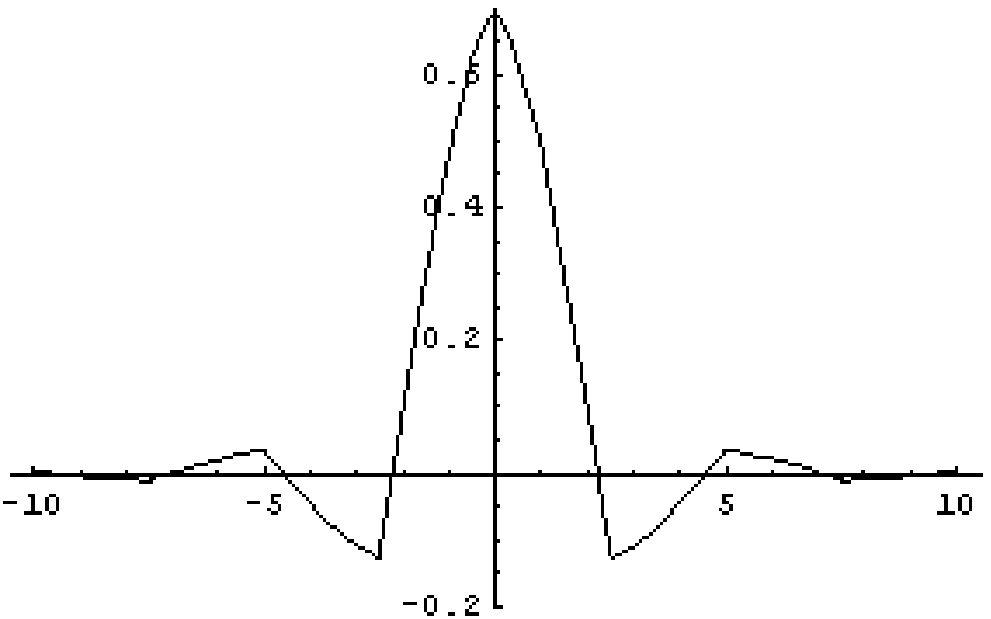}} \hspace{6mm}\hfill\\
\begin{figure}[h]
\caption{\label{fig3} $\Psi(x)$ for $N=0$, $N=1$, $N=2$ (first row)
and $N=3$, $N=4$ (second row) }
\end{figure}
\end{center}

For what concerns the check of the orthonormality, we first notice
that formula (\ref{413}) does not hold here and should be replaced
by the following one, which can be easily analitically derived: \be
<\Psi_{\underline
n}^{(L)},\Psi>=\delta_{n_1,0}\,\sum_{l\in\Bbb{Z}}\,\hat c_l\,\hat
c_{l-n_2}+\frac{1}{\pi(1-4n_1^2L^2)}\,\sum_{l\in\Bbb{Z}}\,\hat
c_l\left(\hat c_{l-n_2+1}+\hat c_{l-n_2-1}\right) \label{413bis}\en
As we see, this formula explicitly depends on $L$, as it should.
Also, replacing the sum on $l\in\Bbb{Z}$ with a finite sum on
$l=-N,\ldots,N$, we can compute the norm of $\Psi(x)$, which already
for $N=4$ is equal to 1 but for an error smaller than $O(10^{-6})$.
An interesting result concerns the overlap between two different
functions $\Psi_{\underline n}^{(L)}$ and $\Psi(x)$. If $n_1=0$ we
find that $\left|<\Psi_{0,n_2}^{(L)},\Psi>\right|\leq 10^{-3}$
already for $N=4$, while if $N=5$ it is even smaller than $5\times
10^{-4}$, $\forall n_2\in\Bbb{Z}$, independently of $L$. However,
$L$ plays a role in the computation of
$\left|<\Psi_{n_1,n_2}^{(L)},\Psi>\right|$ if $n_1\neq 0$. Indeed in
this case we can easily check that, for instance,
$\left|<\Psi_{1,1}^{(L=1)},\Psi>\right|=.155$. This result shows
that, at it was already discussed, our procedure cannot produce an
on basis when $L=1$. A different conclusion is obtained if $L>1$.
Indeed, already for $L=2$, we find that the maximum overlap is given
by $\left|<\Psi_{1,1}^{(L=2)},\Psi>\right|=.031$, already for $N=4$.
This is a reasonable approximation which gets better and better as
$L$ increases: the same quantity equals 0.013 if $L=3$, 0.007 if
$L=4$ and so on. To improve the approximation we should consider a
better approximation for $f(P_1)$. However, such an improvement
necessarily produce a more complicated expression for $c_{\underline
n}^{(L)}$, and will not be considered here.

\vspace{4mm}

More examples are easily constructed starting from functions with
compact support in $[-a,a]$. For these functions the analytic
expressions of $F_L(\underline P)$ can be found with no particular
difficulties, and reasonable approximations can also be deduced in
most of the cases. The situations becomes technically more difficult
 when the starting function $g(x)$ has no compact support, as in
\cite{bagtri}. In this case all the computations are usually more
delicate even if, in principle, they still produce an o.n. basis in
the Hilbert space $\Hil_{g,L}$. More than constructing other
examples we are interested in extending the orthonormalization
procedure to a slightly different situation, i.e. to the case in
which the original set of functions
$\{T_1^{n_1}\,T_2^{n_2}g(x),\,n_1,n_2\in\Bbb{Z}\}$ is constructed
using two unitary operators $T_1$ and $T_2$ which do not commute.
This analysis, in fact, would produce interesting outputs related to
Gabor systems for all of $\Lc^2(\Bbb{R})$ and to systems of
wavelets.

\section*{Acknowledgements}

This work was partially supported by M.U.R.S.T. and partially by the
M.S.R.T. of Iran. MRA wishes to thank the people at the Dipartimento
di Metodi e Modelli Matematici for their hospitality.

\end{document}